\documentclass[10pt, twocolumn]{aastex701}

\usepackage[T1]{fontenc}
\usepackage{mysiunitx}
\usepackage{amsmath}
\usepackage{isomath, natbib, hyperref}
\usepackage{derivative, lipsum, placeins}

\DeclareSIUnit{\Msun}{\mathnormal{M_\odot}}
\DeclareSIUnit{\Rsun}{\mathnormal{R_\odot}}
\DeclareSIUnit{\year}{yr}
\DeclareSIUnit{\erg}{erg}

\newcommand{\Lag}[1]{\ensuremath{{\rm L}_{#1}}}
\newcommand{\teff}[1]{\ensuremath{T_{\rm eff, #1}}}
\newcommand{\gteff}[1]{\ensuremath{\mathcal{T}_{\rm eff, #1}}}

\received{08/13/2025}
\accepted{10/17/2025}

\begin{document}
\title{Attributing the O'Connell effect in contact binaries to a cooling mass-transfer stream}
\author[0000-0003-4200-7852]{Matthias Fabry}
\affiliation{Department of Astrophysics and Planetary Science, 800 E. Lancaster Ave., Villanova, PA 19085, USA}
\email{matthias.fabry@villanova.edu}
\author[0000-0002-1913-0281]{Andrej Pr\v sa}
\affiliation{Department of Astrophysics and Planetary Science, 800 E. Lancaster Ave., Villanova, PA 19085, USA}
\email{andrej.prsa@villanova.edu}

\begin{abstract}
	Contact binaries are very short-period systems that are continuously interacting by transferring mass and energy.
	Obtaining large, statistical samples of contact binaries from photometric surveys can put valuable constraints on the various processes involved in their evolution.
	Modeling those systems however present some challenges.
	In some contact-binary light curves, the O'Connell effect is visible, where the maxima at both quarter phases are unequal.
	In the literature, this effect is typically attributed to magnetic spots on the surface of the binary.
	In this work, we model contact-binary surfaces using PHOEBE, and include a parametric prescription for a lateral mass- and energy-transfer stream that travels from the hotter primary to the cooler secondary.
	We allow this stream to have a variable heat capacity.
	We fit a system from the Kepler sample with a strong O'Connell effect, and show that a low-heat capacity stream can explain the unequal maxima.
	This suggests that, in such systems, surface flows can play a significant role in transferring heat between components.
	Our methods can be used on larger samples of contact binaries from OGLE, Kepler, or TESS to advance our understanding of contact binary structure and evolution.
\end{abstract}

\section{Introduction}
Contact binaries, also called W UMa stars after their prototype, have been observed since variable-star catalogs became available \citep[e.g.,][]{eggenContactBinariesII1967, binnendijkOrbitalElementsUrsae1970}.
Their light curves (LCs) have short periods ($p \lesssim \qty{0.5}{\day}$), are smoothly varying, and are characterized by wide eclipses.
The eclipse depths are also very similar, leading to the inferred temperature ratio being close to unity, independent of the mass ratio of the system.
Because of the proximity of the components, the inclination of contact binaries can be quite low while still being partially eclipsing.

Contact-binary LCs are widely modeled in the literature using a Wilson-Devinney method \citep{wilsonRealizationAccurateCloseBinary1971}.
PHOEBE \citep{prsaPhysicsEclipsingBinaries2016,horvatPhysicsEclipsingBinaries2018,jonesPhysicsEclipsingBinaries2020, conroyPhysicsEclipsingBinaries2020} is a modern implementation of this method, and includes various secondary effects like mutual irradiation, different limb-darkening laws and relativistic boosting.
It also allows accurate modeling of contact-binary envelopes, by using a triangular marching-mesh method that smoothly connects the two parts in the neck region \citep{prsaPhysicsEclipsingBinaries2016}.
However, up to PHOEBE 2.4, the two halves of the envelope still carry separate surface properties, in particular effective temperature.
The halves are thus thermally isolated, while nearly all W UMa systems are observed in thermodynamical equilibrium.

Contact binaries are ubiquitous, leading to the conclusion that they must form from short-period, main-sequence binaries, and that they persist over the nuclear timescale of the component stars.
No complete model of contact-binary evolution exists yet.
It is thought that the effects of mass transfer (MT), energy transfer (ET), tidal friction from magnetic braking, and von Zeipel-Kozai-Lidov oscillations due to third bodies are important ingredients \citep[e.g.,][]{lucyLightCurvesUrsae1968, flanneryCyclicThermalInstability1976, shuStructureContactBinaries1979, webbinkEvolutionLowmassClose1977, kahlerStructureContactBinaries2004, liStructureEvolutionLowmass2005, stepienLargescaleCirculationsEnergy2009, eggletonFormationEvolutionContact2012}.

The endpoint of contact-binary evolution is the merger of the two bodies into one, which is initiated either by the tidal Darwin instability at a certain limiting mass ratio \citep{darwinDeterminationSecularEffects1879,rasioMinimumMassRatio1995,liDynamicalStabilityUrsae2006, pestaMassratioDistributionContact2023}, or by the unstable mass loss from the outer Lagrangian point \citep{shuStructureContactBinaries1979, pejchaBuryingBinaryDynamical2014, pejchaCoolLuminousTransients2016}.
Stellar mergers are accompanied with transient events called Luminous Red Novae, and V1309 Sco has been confirmed as the merger of a contact binary \citep{tylendaV1309ScorpiiMerger2011}.

Although not exclusive to contact binaries, in some LCs, one maximum is brighter than the other.
This is the so-called O'Connell effect, after \citet{oconnellSocalledPeriastronEffect1951} showed that the periastron explanation from \citet{robertsMethodDeterminingAbsolute1906} and \citet{duganEclipsingVariablesRV1916} was incorrect.
\citet{wilseyRevisitingOConnellEffect2009} state that modeling the O'Connell effect with dark star spots has been the most popular so far.
However, some authors \citep[see references in][]{wilseyRevisitingOConnellEffect2009} made the ad-hoc choice of modeling their LCs with a single, large spot that covers a significant fraction of the stellar surface, which seems nonphysical under the interpretation of chromospheric activity.
As an alternative to star spots, \citet{zhouExplanationLightCurveAsymmetries1990} explained the asymmetry by internal circulation currents that are deflected by the Coriolis force, resulting in a hotter and a cooler stream on opposite sides of a contact binary.
This hypothesis could explain those systems where the higher maximum follows the secondary minimum \citep{mccartneyStatisticalStudyOconnell1997}, but is rather unlikely for the inverse case, as the streams have to be counter-rotating in the orbital frame of reference.
A further model to explain the O'Connell effect was put forth by \citet{liuPossibleExplanationOConnell2003}, in which the unequal brightness of the components is explained by the thermalization of circumbinary material that impacts the stars.
In \citet{stepienLargescaleCirculationsEnergy2009}, this hypothesis was put on firmer grounds, by the introduction of a hydrodynamic model of energy transfer in general contact-binary configurations.
As yet another hypothesis, stream impacts have been proposed to explain the unequal maxima.
Rather than dark spots, this model features a hot spot where a mass-transfer stream impacts the secondary.
Recently, \citet{rucinskiLessonsHighresolutionSpectroscopy2025} found spectroscopic evidence of stream impacts on the contact binaries AW UMa and $\epsilon$ CrA.

In order to put constraints on the ET process, we propose a model that explains contact-binary LCs with a strong O'Connell effect by including a MT stream encompassing the cooler secondary star.
This paper is structured as follows.
In Sect.~\ref{sec:methods}, we outline our model, while in Sect.~\ref{sec:kic} we apply it a system observed by the Kepler mission.
Section \ref{sec:conc} discusses the results and closes with some concluding remarks.

e\section{Proposed model}\label{sec:methods}
In this section, we detail two models of ET that adjust the temperature of the surface of the secondary star as a function of the properties of both components.
We use a right-handed coordinate system $\vec{r} = (x, y, z)$, with the origin at the center of mass, the $x$-axis connecting the two stars, and $z$ pointing in the direction of total angular momentum.
Let $\gteff{i}$ be the global, surface-averaged effective temperature of component $i$ and $\teff{i}(\vec{r})$ the local effective temperature for each surface element of component $i$.
Non-primed temperatures refer to values before the ET model is applied, and primed after ET.

In the first model of ET, we assume very efficient thermal contact, so that the entire envelope of the binary has one global $\mathcal{T}_{\rm eff}$.
This is expected if the internal structure of the envelope is shellular and barotropic \citep{shuStructureContactBinaries1976,fabryModelingContactBinaries2023}.
While we recognize that this model may only be valid for systems with radiative envelopes evolving on the nuclear timescale \citep{hazlehurstStabilityAgezeroContact1980, hazlehurstEquilibriumContactBinary1993, kahlerStructureContactBinaries2004, fabryModelingContactBinaries2025}, it is still useful to note here as it is parameter free.
Each surface element of the secondary is then scaled to the (global) temperature ratio,
\begin{equation}
	\teff{2}'(\vec{r}) = \frac{\gteff{1}}{\gteff{2}}\teff{2}(\vec{r}).
\end{equation}
This process, which we call ``internal mixing,'' reduces the dimensionality of the LC-fitting problem by one (by eliminating $\gteff{2}$), but it has no asymmetries.
As such, while it can explain the equal depths of most W UMa LCs, it cannot explain the unequal maxima in those with the O'Connell effect.

A second way of modeling ET is by considering a MT stream from the hotter to the cooler component.
In systems in shallow contact, where the internal structure is still highly baroclinic, surface flows are expected to play a role in equalizing the temperatures of the components \citep{zhouExplanationLightCurveAsymmetries1990, liuPossibleExplanationOConnell2003, kahlerStructureContactBinaries2004}.
\citet{stepienLargescaleCirculationsEnergy2009} calculated a model with mass and energy flows that partly enshrouds the cooler secondary.
He showed that, for the system AB And, the heat capacity of the stream is high, meaning that it hardly cools while traversing the secondary, but the theory permits thinner streams with more limited heat capacities.
In our model, we allow the stream to cool significantly and blend into the envelope of the secondary, as opposed to sinking into the primary when it has completed the lap around the secondary.
Such ET can result in a steady-state profile where the temperature decreases longitudinally.

\begin{figure*}  
	\centering
	\includegraphics[width=0.9\textwidth]{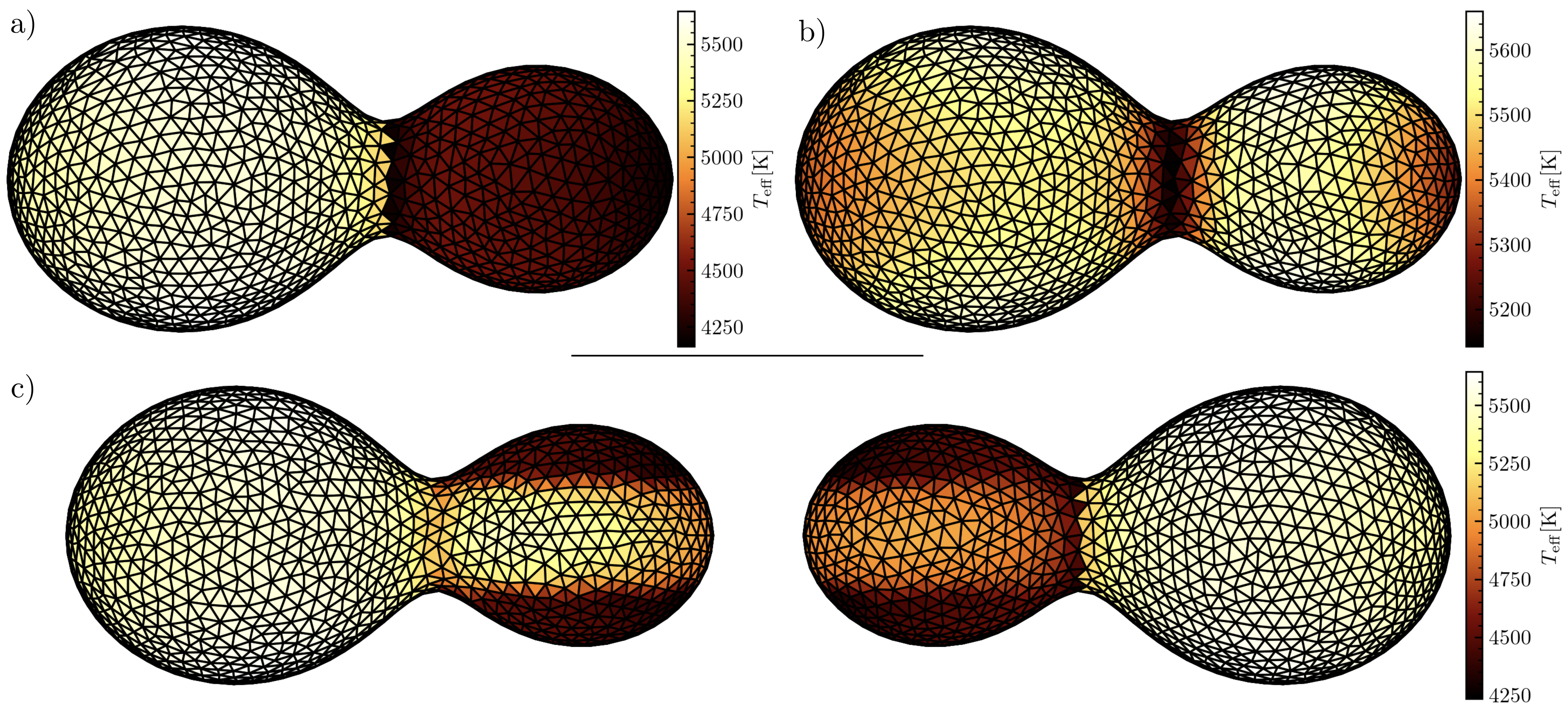}
	\caption{PHOEBE meshes of different energy transfer modes where parameters $\gteff{1} = \qty{5500}{\kelvin}$, $\gteff{2} = \qty{4400}{\kelvin}$, $F = 0.5$ and $q = 0.5$ are used throughout. Panel (a): no mixing applied. Panel (b): internal mixing mode. Panel (c): Two quadratures of the lateral mixing mode. Note the different color scale in panel b.}
	\label{fig:meshes}
\end{figure*}

We model such an ET stream on the surface of a contact binary as follows.
First we set the latitudinal extent of the steam, $z_{\rm stream}$, which can vary, in principle, from zero to the total polar radius of the secondary.
In practice, however, as modeled by \citet{stepienLargescaleCirculationsEnergy2009}, stream extents will be around $\qty{15}{\degree}$ to $\qty{30}{\degree}$ as measured on a meridian on the secondary, corresponding to fractional $z$-heights of around 0.25 to 0.5.
Then, for any surface element where $|z| < z_{\rm stream}$, we adjust its temperature according to
\begin{equation}
	\teff{2}'(\vec{r}) = \teff{2}(\vec{r})\left(1+\left(\frac{\gteff{1}}{\gteff{2}}-1\right)f(z)g(\phi)\right).
\end{equation}
The functions $f$ and $g$ represent the latitudinal and longitudinal dependence, respectively, of the ET process:
\begin{align}
	f(z) &= \left(1 - \frac{|z|}{z_{\rm stream}}\right) ^ p, \\
	g(\phi) &= \left(\frac{\phi_{\rm max} - \phi}{\phi_{\rm max} - \phi_{\rm min}}\right) ^ s,
\end{align}
with $\phi = \arctan(y/x) + \pi$ the longitudinal angle of the MT stream on the surface of the secondary, and $p$ and $s$ are what we call the latitudinal and longitudinal shape parameters, respectively.
This mode of ET we call ``lateral transfer.''
With $s \ne 0$, this mode is asymmetrical at opposite quadratures.
Conversely, with $z_{\rm stream}$ equal to the polar radius and $p = s = 0$, this model reduces to the internal mixing mode above.

Comparing to the hydrodynamic model of \citet{stepienLargescaleCirculationsEnergy2009}, we note that the parameter $s$ loosely maps to the heat capacity of the stream.
The higher the heat capacity, the lower $s$ will be, and the slower the stream cools as it travels around the secondary.
The height of the stream, $z_{\rm stream}$ is determined by the meridional pressure balance, while the parameter $p$ qualitatively defines how quick heat is lost in the latitudinal direction (once again a proxy for heat capacity).

\begin{figure}
	\centering
	\includegraphics[width=\columnwidth]{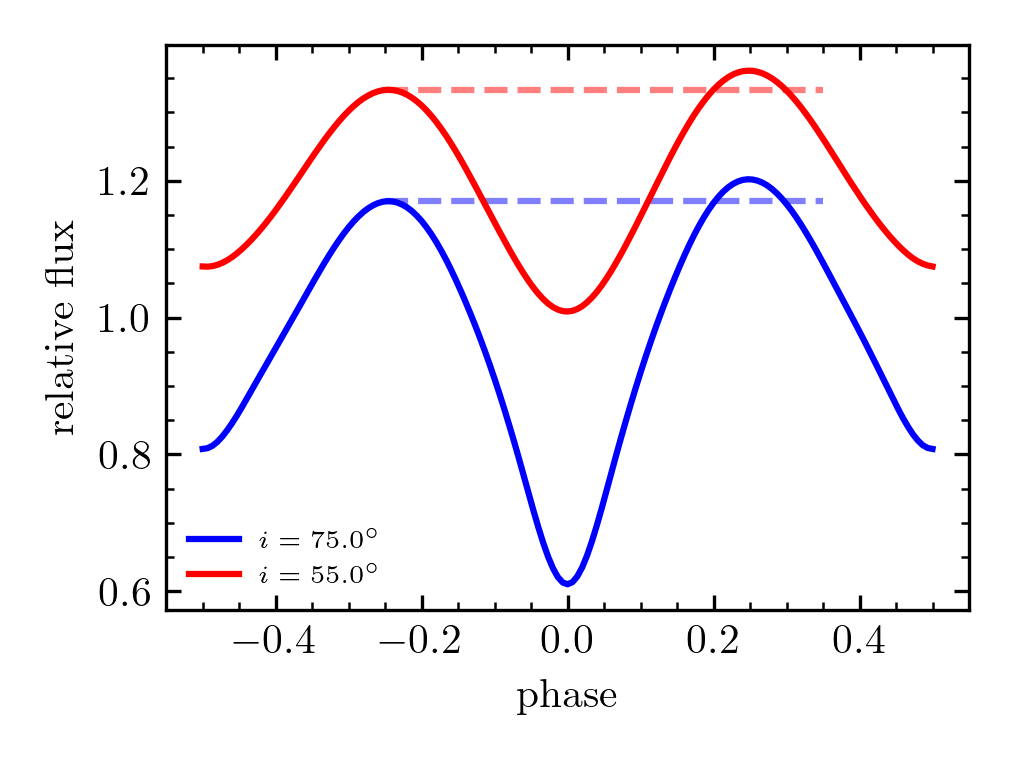}
	\caption{Light curves of the same binary system with lateral mixing of Fig.~\ref{fig:meshes}, but viewed under different inclinations. The O'Connell effect is highlighted with dashed lines. The unit of the $y$ axis is arbitrary.}
	\label{fig:inclination}
\end{figure}
Figure \ref{fig:meshes} depicts PHOEBE meshes of a contact binary with temperatures $\gteff{1} = \qty{5500}{\kelvin}$ and $\gteff{2} = \qty{4400}{\kelvin}$, mass ratio $q = 0.5$, filling factor $F = (\Psi_{\Lag{2}} - \Psi)/(\Psi_{\Lag{2}} - \Psi_{\Lag{1}}) = 0.5$ and $p = s = 0.2$.
Panel (a) shows the mesh without ET, panel (b) with internal mixing, while panel (c) shows both quadratures of the lateral mixing mode.
Comparing the meshes reveals that the lateral mixing models also introduce different eclipse depths due to the presence of the cooler poles.
This ratio will be sensitive to the inclination and the width of the MT stream.
The LC of the mesh in panel (c) viewed under different inclinations is shown in Fig.~\ref{fig:inclination}, where it is apparent that under higher inclinations, the cooler poles of the secondary can totally eclipse the hot primary and produce a deeper minimum than under lower inclinations, where the hot poles of primary are always visible.

\section{Application to KIC 6223646}\label{sec:kic}
As a proof of concept, we apply the lateral ET mode of Sect.~\ref{sec:methods} to a system observed by the Kepler mission \citep{boruckiKeplerPlanetDetectionMission2010}.
The data used here is the detrended LC from the eclipsing binary catalog from \citet{kirkKeplerEclipsingBinary2016}\footnote{\url{https://keplerebs.villanova.edu}}.
We choose the system KIC 6223646 for two reasons.
It is a 13th magnitude system and thus has very high-quality data, and its LC features one of the strongest O'Connell effects in the Kepler sample, with difference of maxima over 5\%.

For the modeling, we assume a fixed period $P = \qty{0.36493416}{\day}$ and periastron passage epoch $T_0 = \num{54953.71534601} ({\rm BJD - 2400000})$, taken from \citet{kirkKeplerEclipsingBinary2016}.
From the analysis of the LAMOST spectrum, \citet{frascaActivityIndicatorsStellar2016} derived an effective temperature of $\mathcal{T}_{\rm eff} = \qty{9685}{\kelvin}$ and a spectral type of B9III.
Given the binary nature of this source, the used pipeline is likely not accurate for this system.
Taking the period-\teff{} relation from \citet{jayasingheASASSNCatalogueVariable2020}, a period of $P=\qty{0.364}{\day}$ should correspond to either $\qty{5878}{\K}$ or $\qty{7157}{\K}$ if it is a late-type or early-type contact binary, respectively.
Furthermore, it is difficult to fit a late B/early A-type star in a contact configuration without overflowing the \Lag{2} equipotential.
Nevertheless, as the LC is largely independent of the primary temperature, we use $\gteff{1} = \qty{9000}{\kelvin}$ in the photometric modeling.
The maximum-likelihood parameters and $1\sigma$ uncertainties obtained from \texttt{emcee} \citep{foreman-mackeyEmceeMCMCHammer2013} are presented in Table \ref{tab:par6223646}.
The corner plot showing the posterior distribution is depicted in Fig.~\ref{fig:corner} in App.~\ref{app}.
The errors quoted in Table \ref{tab:par6223646} are formal errors only, and are underestimated because model uncertainties are not taken into account.
We refer to Sect.~\ref{ssec:modeluncer} below for a more detailed discussion and a strategy to estimate their magnitude.

\begin{figure}
	\centering
	\includegraphics[width=\columnwidth]{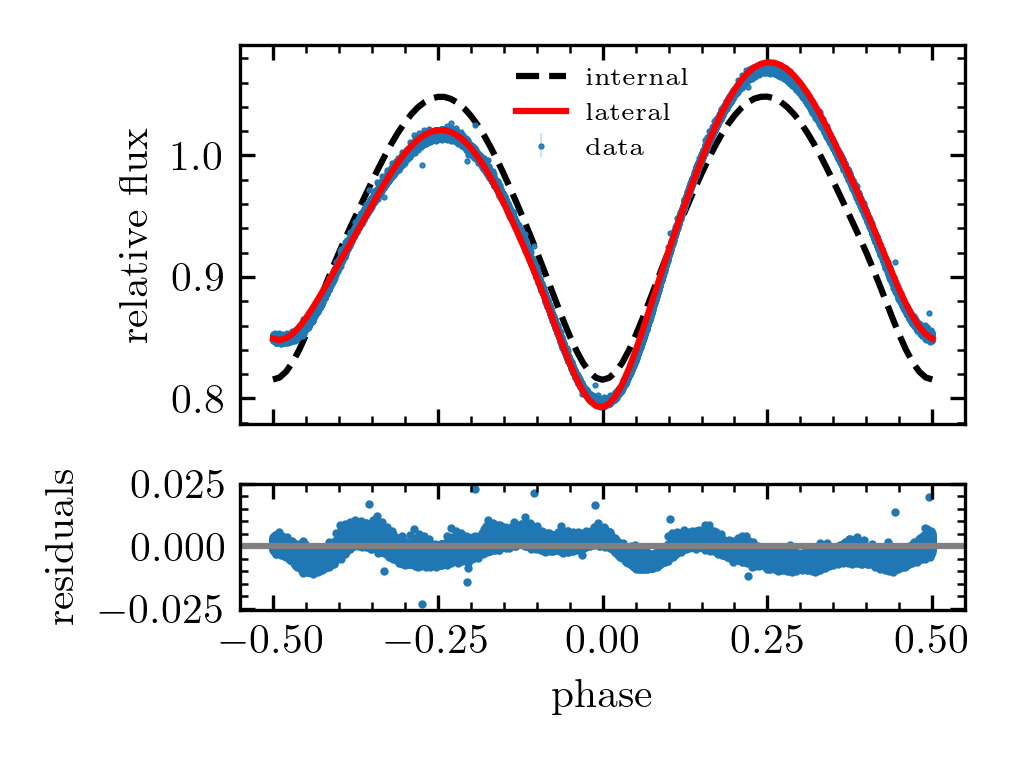}
	\caption{Light curve for KIC 6223646 (blue), with the best-fit PHOEBE model with lateral mixing in red. The lower panel shows the residuals, which has a root-mean-squared value of 0.0035. For comparison, an equal-temperature model (by internal mixing) is shown in black.}
	\label{fig:lc6223646}
\end{figure}

Figure \ref{fig:lc6223646} shows the LC along with the best-fit model.
For comparison, an internal mixing model is overplotted in dashed lines, which cannot reproduce either the different eclipse depths or the O'Connell effect.
Note that it also cannot explain the slight asymmetric ingress/egress of the eclipses.

The corner plot (Fig.~\ref{fig:corner}) shows that the mass ratio, $q$, is anti-correlated with $\gteff{2}$, the longitudinal cooling rate $s$, and, to a lesser extent, the filling factor $F$ and inclination.
This can be understood from geometrical considerations.
At lower mass ratios, the fractional area of the secondary (with respect to the whole system) is lower, so its surface temperature needs to increase to match the minima, while the ET needs a steeper drop-off by increasing $s$ (because the LC still needs to match the unequal maxima of the O'Connell effect).
An increase in $F$ raises the overall projected area of the secondary, while the increase in inclination ensures the components produce the unequal eclipse depths.
The parameters $p$ and $z_{\rm stream}$ are, expectedly, strongly correlated over a wide range, suggesting that, in future fits, considering a fixed latitudinal shape $p$ could be chosen instead.

\begin{table}
	\centering
	\caption{Maximum-likelihood parameters for the lateral-mixing model of KIC 6223646. Quoted uncertainties are formal 1-$\sigma$ errors from MCMC estimation.}
	\begin{tabular}{r|l}
		Parameter (Unit) & {Value}\\  \hline\hline
		$q$ & $\num{0.461(25:27)}$\\
		$i$ (deg) & $\num{60.24(27:26)}$\\
		$F$ & $\num{0.244(19:17)}$\\ \hline
		$\gteff{1} (\unit{\kelvin})$ & $\num{9000}$ (Fixed) \\
		$\gteff{2} (\unit{\kelvin})$ & $\num{6990(130:130)}$\\
		$p$ & $\num{0.88(45:52)}$\\
		$s$ & $\num{1.06(19:17)}$\\
		$z_{\rm stream}$ & $\num{0.72(19:25)}$
		
	\end{tabular}
	\label{tab:par6223646}
\end{table}

For this system, we find that the temperatures, which are commonly assumed equal for W UMa systems, are substantially different if we account for a MT stream around the secondary.
Instead of the secondary being roughly $\qty{9000}{\kelvin}$ (barring the potential mis-classification of the primary, as discussed above), the effective temperature is closer to $\qty{7000}{\kelvin}$, a 20\% difference.
In some cases with early-type primaries, this can mean that the temperature of the secondary can drop below the radiative-convective envelope transition around $\qty{6500}{\kelvin}$, which will change the expected evolution of the system.

The mass ratio being around $0.45$ and filling factor $F=0.24$ (which corresponds roughly to $R/R_{\rm RL} \approx 1.07$) is consistent with this binary undergoing contact evolution with mass and energy transfer.
If both stars in KIC 6223646 have radiative envelopes, which would be the case if the $\gteff{1}$ from the LAMOST analysis is to be assumed, then we expect the system to evolve as computed in \citet{menonDetailedEvolutionaryModels2021} and \citet{fabryModelingContactBinaries2025}, toward equal masses.
However, as argued above, it is unlikely that an early A-type star can be in a stable contact configuration with an F-type companion at the observed orbital period and computed overflow.
Therefore, if we assume the system contains lower-mass components, e.g., an F-type with a G-type, at least one star will have a deep convective envelope and the evolution is expected to proceed toward more unequal masses (e.g., \citealt{flanneryCyclicThermalInstability1976}, \citealt{lucyUrsaeMajorisSystems1976}, \citealt{yakutEvolutionCloseBinary2005}, Fabry \& Pr\v sa 2025b, in prep.).
The different evolutionary scenarios could be distinguished by measuring the period derivative, although all sources of angular momentum loss (such as from magnetic braking or third bodies) would also needed to be taken into account.

\subsection{Model uncertainties}\label{ssec:modeluncer}
We note that the residuals of the fit in Fig.~\ref{fig:lc6223646} do not represent Gaussian noise.
This signals that there are uncertainties in the overall error budget we did not account for.

Unfortunately, gauging the uncertainties of the model is not straightforward, as it amounts to estimating unknown parameters that have not been fitted, e.g., spots, a more complex geometry of the MT stream or departure from the Roche model \citep[see, e.g.,][]{rucinskiLessonsHighresolutionSpectroscopy2025}.
Given the current parameters of the model however, we can estimate how wide we should take the uncertainty distributions by comparing the family of LCs computed with parameters from those distributions with the observations.
To do this, we first convert the posteriors to a multi-variate Gaussian distribution (by resampling \num{e6} times), and obtain its covariance matrix $\Sigma_{k\ell}$.
Next, for a parameter that appears at index $k$ in the matrix, we scale its entries with a factor $\sqrt{\kappa_k}$:
\begin{subequations}
	\begin{align}
		\Sigma_{k\ell} &\rightarrow \sqrt{\kappa_k}\Sigma_{k\ell},\\
		\Sigma_{\ell k} &\rightarrow \sqrt{\kappa_k}\Sigma_{\ell k}.
	\end{align}
\end{subequations}
Diagonal entries, $\Sigma_{kk}$ are thus multiplied in total by $\kappa_k$.
Note that, because our posteriors are not Gaussian and correlated, such scalings cannot be interpreted simply as the factor with which the parameter uncertainties are underestimates.
They will give us a qualitative idea however which parameters have the most impact on the remaining variance in the data.
We then resample the distributions with the scaled covariance, and compute the 1-$\sigma$ envelope of model flux in each phase bin.
Because the parameters are highly correlated, there is no straightforward way to search for the factors $\kappa_k$.
Finally, using the scaled covariance matrix, we take 100 samples from the distribution and compute the model fluxes.
In each of 50 phase bins, we determine the median and the 1-$\sigma$ flux values.

\begin{figure}
	\centering
	\includegraphics[width=\columnwidth]{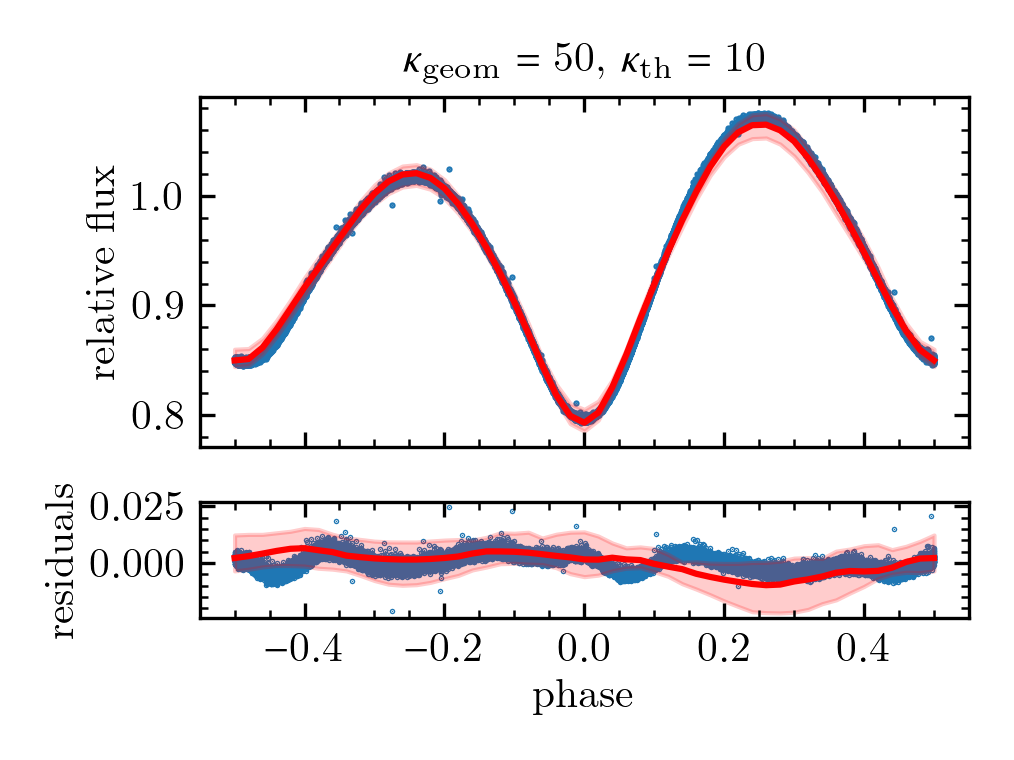}
	\caption{PHOEBE models of KIC 6223646 that include larger covariance than the original MCMC estimate in Table \ref{tab:par6223646}. The red line is the median of the 200 models from the rescaled distributions, while the shaded area is the 1-$\sigma$ flux range of these models.}
	\label{fig:errorestimate}
\end{figure}

Finding a global minimum in $\kappa_k$ space where the 1-$\sigma$ flux values best approximate the variation in the data would require another minimization and MCMC exploration, which is beyond the scope of this work.
Instead, we progressively increase $\kappa_k$ until the root-mean-squared values of the 1-$\sigma$ flux values approaches 0.01.
We group the geometrical parameters, $q$, $i$ and $F$, and refer to to their scalings collectively refer as $\kappa_{\rm geom}$, while we set $\kappa_{\gteff{2}} = \kappa_r = \kappa_p = \kappa_{z_{\rm stream}} = 1$ and collectively call them $\kappa_{\rm th}$.
With this procedure, we find $\kappa_{\rm geom} = 50$ and $\kappa_{\rm th} = 10$, and the LC fit is shown in Fig.~\ref{fig:errorestimate}.
From the reconstructed, scaled multivariate Gaussian posteriors (see Fig.~\ref{fig:wide_corner}), we estimate that the errors on the geometrical parameters are underestimated by at least a factor 7.
The scaled posteriors show that the thermal parameters have their 1-$\sigma$ interval increased by around a factor of 3.
A more rigorous error analysis is needed, for example using a grid of models and employing the Mahalanobis distance (see, e.g., \citealp{johnsonAppliedMultivariateStatistical2002}), could be used to obtain more accurate parameter uncertainties.

\section{Conclusions}\label{sec:conc}
In this work, we introduced a method to model contact binary LCs with an O'Connell effect by including a cooling MT stream from the hotter to the cooler component.
By adjusting the temperature of the surface elements of the cooler secondary with a longitudinally dependent function, asymmetries were introduced that can explain the unequal maxima of the O'Connell effect, which is observed in various contact-binary LCs.
Such a stream is supported by the theory of \citet{stepienLargescaleCirculationsEnergy2009} when the stream is thin and has a low heat capacity.
A notable shortcoming of our method relates to model uncertainties, which are apparent in the fit of KIC 6223646.
Scaling the covariance matrix of the MCMC sampling revealed that the uncertainties of the geometrical parameters of mass ratio, filling factor and inclination are likely underestimated by an order of magnitude.
These errors are to be revised in further work.
\par

Applying a mixing scheme akin to one developed here over a large sample of contact-binary LCs, such as the Kepler or TESS eclipsing binary catalogs \citep{kirkKeplerEclipsingBinary2016, prsaTESSEclipsingBinary2022}, can provide valuable insight into the ET processes.
In particular, examining the correlation between secondary temperatures and the filling factor may provide evidence that, for shallow contact binaries, ET is surface-flow dominated, while for the systems in deep contact, internal mixing dominates.
On the other hand, according to \citet{stepienLargescaleCirculationsEnergy2009}, constraining the latitudinal extent of the stream (parameterized with $z_{\rm stream}$ in this work) can provide insight into the thermal properties of the MT stream.
Such constraints may prove valuable for producing evolutionary models of W UMa binaries, in which both the efficiency and location of the ET process are free parameters.

\begin{acknowledgments}
	F.M.~and P.~A.~recognize the support from the National Science Foundation under grant no.~AST-2306996.
\end{acknowledgments}

\bibliography{massive_stars}
\bibliographystyle{aasjournal}

\appendix
\FloatBarrier

\section{Additional figures}\label{app}
In this appendix, we provide additional figures that were deferred to improve readability of the main text.

\begin{figure*}
	\includegraphics[width=\textwidth]{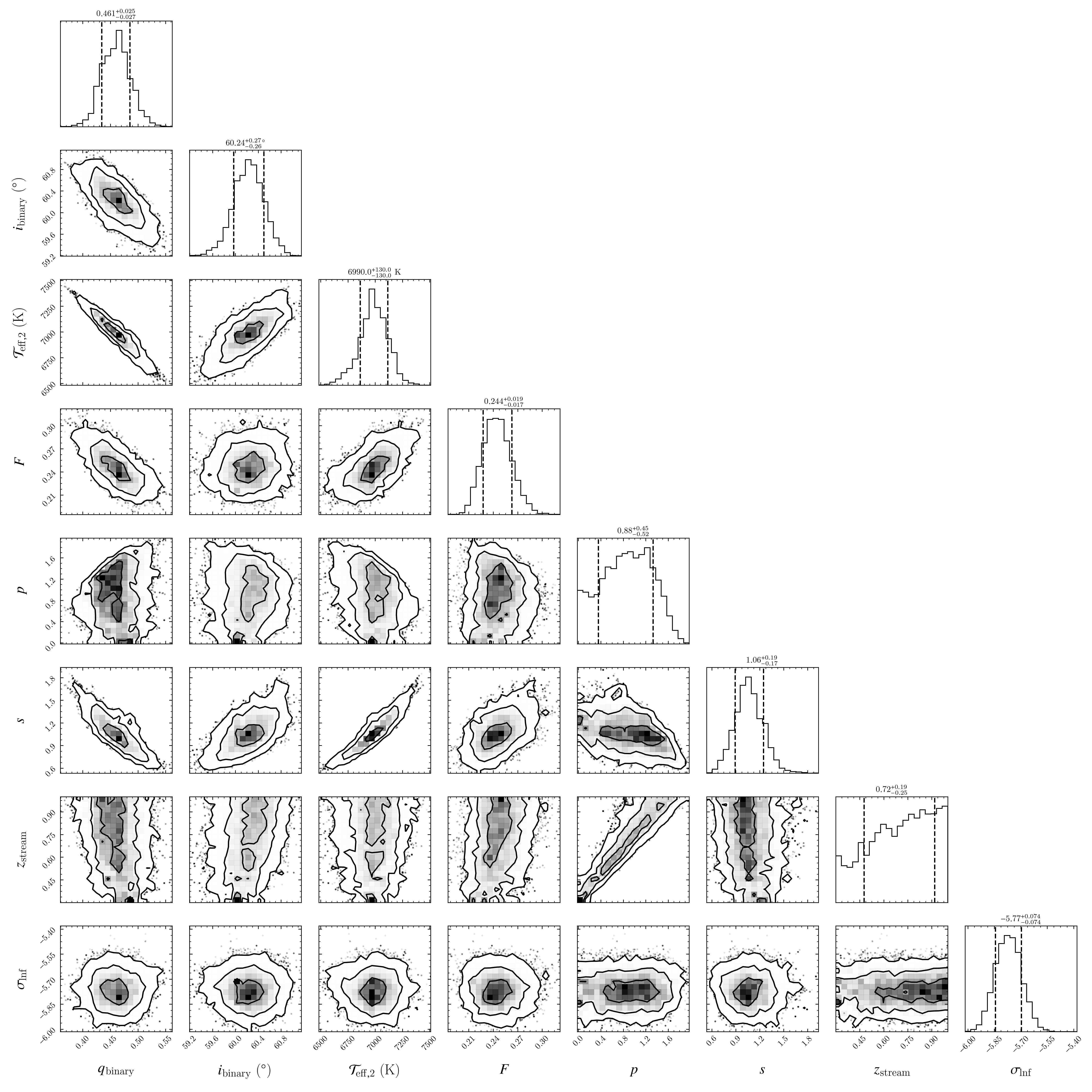}
	\caption{Posterior distribution densities of the MCMC chains that estimated the errors on the fit parameters in Table \ref{tab:par6223646}. The final row, $\sigma_{\rm lnf}$ is a numerical ``fudge'' parameter that estimates the fraction with which the errors on the data are underestimated.}
	\label{fig:corner}
\end{figure*}

\begin{figure}
	\centering
	\includegraphics[width=\textwidth]{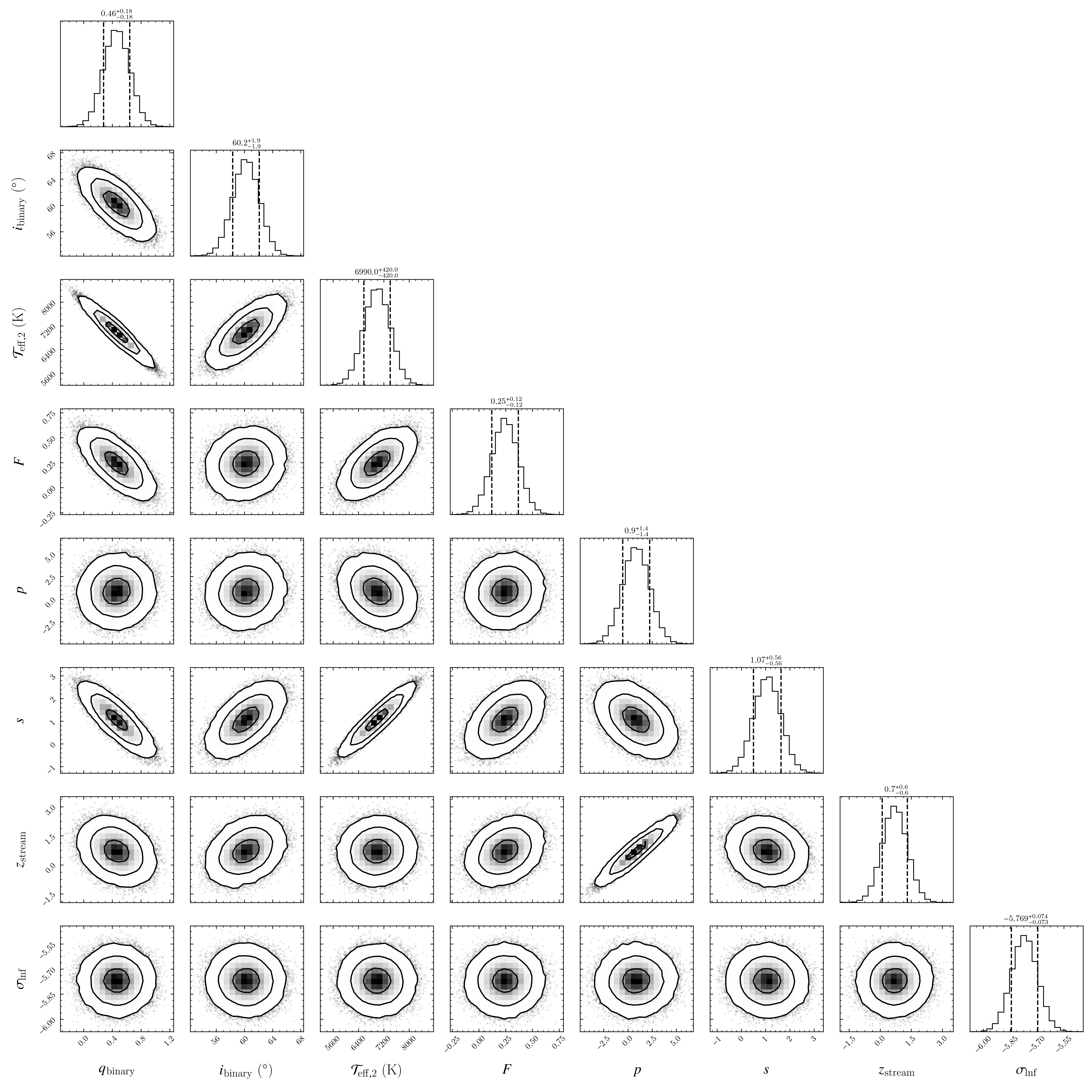}
	\caption{Corner plot of the reconstructed, scaled multivariate Gaussian posteriors. }
	\label{fig:wide_corner}
\end{figure}

\end{document}